\documentclass[sigconf]{aamas} 

\usepackage{balance} 

\setcopyright{ifaamas}
\acmConference[AAMAS '21]{Proc.\@ of the 20th International Conference on Autonomous Agents and Multiagent Systems (AAMAS 2021)}{May 3--7, 2021}{Online}{U.~Endriss, A.~Now\'{e}, F.~Dignum, A.~Lomuscio (eds.)}
\copyrightyear{2021}
\acmYear{2021}
\acmDOI{}
\acmPrice{}
\acmISBN{}

\acmSubmissionID{???}

\title[AAMAS-2021 Zhao]{Mechanism Design Powered by Social Interactions}
\subtitle{Blue Sky Ideas Track}

\author{Dengji Zhao}
\affiliation{
  \institution{ShanghaiTech University}
  \city{Shanghai, China}
  }
\email{zhaodj@shanghaitech.edu.cn}

\begin{abstract}
Mechanism design has traditionally assumed that the set of participants are fixed and known to the mechanism (the market owner) in advance. However, in practice, the market owner can only directly reach a small number of participants (her neighbours). Hence the owner often needs costly promotions to recruit more participants in order to get desirable outcomes such as social welfare or revenue maximization. In this paper, we propose to incentivize existing participants to invite their neighbours to attract more participants. However, they would not invite each other if they are competitors. We discuss how to utilize the conflict of interest between the participants to incentivize them to invite each other to form larger markets. We will highlight the early solutions and open the floor for discussing the fundamental open questions in the settings of auctions, coalitional games, matching and voting. 
\end{abstract}

\keywords{Mechanism Design; Social Networks; Diffusion Incentives; Auctions; Coalitional Games; Matching; Social Choice}

\newcommand{\BibTeX}{\rm B\kern-.05em{\sc i\kern-.025em b}\kern-.08em\TeX}

\begin{document}


\pagestyle{fancy}
\fancyhead{}


\maketitle 


\section{Introduction}
Mechanism Design studies how to implement desirable social choice functions in a strategic environment where all participants act rationally in game theoretical sense. Auction as one of the key outputs of mechanism design has been widely used in different markets for a long history (dates back to 17th century)~\cite{shubik2004theory}. Vickrey auction is the seminal work of Vickery~\cite{vickrey1961counterspeculation} which inspired many auction theories such as Vickrey-Clarke-Groves (VCG) auction~\cite{Clarke1971,Groves1973}, Gibbard-Satterthwaite theorem~\cite{gibbard1973manipulation,satterthwaite1975strategy}, Myerson's revenue-maximizing auction~\cite{myerson1981optimal} and Myerson-Satterthwaite theorem~\cite{myerson1983efficient}. In addition to the rich theoretical results, in the beginning of 21st century, IT service providers like Google started to apply a modification of Vickrey auction, called generalized second-price (GSP) auction, to allocate ad impressions~\cite{GSP2007}. Although GSP does not have the desirable property called truthfulness as Vickrey auction does, it has been a golden mechanism for online advertising. 

As the technology of smart devices keeps pushing the boundaries in the last decade, more and more IT services have been moved to smart devices such as online shopping and online games. Many social networking applications like TikTok are purely built on smart devices. This also pushed the online advertising market shifting from traditional PC-based channels to smart-device-based channels~\cite{RDF_W3}. 
More importantly, we see many new forms of mechanisms for online advertising/shopping 
utilising users' connections/interactions on social networks~\cite{TikTok,TikTok2}. 

As the Internet is moving from traditional PC-network to social networks and in the near future to Internet of Things,
we have more personal information from the Internet than before. For example, from a smart phone, with its owner's permission, we can get the owner's social connections, preferences, locations, photos, reviews, shopping history and etc.. This opens up a huge new space for market design which, of course, is not just for advertising.

In a mechanism design setting, we model all the information that a mechanism needs to elicit from the participants as their types. In the literature, the types are typically cardinal or ordinal preferences on outcomes. Moreover, the participants are mostly assumed to be independent. However, in the modern economy underpinned by social networks, people are well-connected. They can quickly gather together online to share resources, distribute tasks or make decisions, even though they are not physically together. Hence, it is essential to explicitly model and utilize their connections in the corresponding market design stage.


In this paper, we propose to utilize people's social connections to invite each other to build larger markets, which enables the market to achieve better outcomes. However, people would not invite each other when there is a competition among them, say, in a limited resource allocation. We discuss how to resolve the conflict of interest among the participants to incentivize them to invite each other, especially, in the settings of resource allocation, task allocation, matching and voting.

In the four mentioned settings, it is easy to see both the benefits and challenges of attracting more participants. 
\begin{itemize}
\item In resource allocation (auctions), a larger market will discover more participants' valuations/demand and increase social welfare or the seller's revenue. The challenge is to ask participants to invite other participants to compete for the same resources.
\item In task allocation (coalitional games), a larger group of participants creates larger coalitions (better outcomes/utilities). 
For example, in a research project, it is always good to add someone with different skills to the team. 
However, the newly added member might also have some skills which the team already has, which creates a competition for the reward sharing among the participants. 
\item In matching, a larger group of participants makes more satisfiable matchings, e.g., larger exchange cycles. The same as in the resource allocation, newly invited participant may compete with their inviters for the match. 
\item In voting, when we have more voters to vote, it will not only increase the turnout rate but also make the voting results harder to manipulate. 
The challenge is that a voter would not invite someone with different preferences.
\end{itemize}
In the rest of the paper, we will model the challenges and discuss possible techniques to incentivize participants to invite their competitors in the four settings. Although they share a similar challenge, the methods to tackle it are very different. Other mechanism design settings such as facility allocation and public goods are not discussed here, but they all can be studied under this framework.


\section{The General Model}
Here we describe a general model of mechanism design on social networks. We consider a game with $n$ players and they are connected via their social connections to form a connected social network. Let $N$ be the set of all players. Each player $i\in N$ has a type $\theta_i = (r_i, p_i)$, where $r_i \subset N$ is $i$'s neighbours (with whom $i$ can directly communicate and $i$ does not know the others $N \setminus r_i$) in the network and $p_i$ is $i$'s other private information defined for the specific game. For example, in a single item auction, $p_i$ is $i$'s valuation for the item. In a house allocation, $p_i$ can be $i$'s preference on all exchangeable houses. In all the different settings, one common parameter of $i$'s type is her neighbours $r_i$.

To execute a mechanism in the model, we need a mechanism/market owner. In this model, the owner can be a special player in the network. For example, a seller for an item or a sponsor for a set of tasks. Let ${\omega}\in N$ be the mechanism owner. It is evident that without attracting more participants, $\omega$ can only run the mechanism among her neighbours $r_{\omega}$. In addition to the traditional goals, the new goal of the mechanism design here is to incentivize $r_{\omega}$ to invite their neighbours to join the mechanism and the newly invited players would further do the same. Eventually, everyone from the network is invited and the owner can run the mechanism among all of them.

In the traditional mechanism design settings, $i$'s type is just $p_i$. One important property called \emph{incentive compatibility} (a.k.a. truthfulness) is defined as "if a mechanism is incentive compatible, then for all $i\in N\setminus \{\omega\}$, reporting $p_i$ truthfully to the mechanism is a dominant strategy". In our model, the type space is enlarged and we need to extend the definition of incentive compatibility to cover the action of inviting their neighbours. To model the invitation action mathematically, we also ask them to report their neighbour sets to the mechanism, which does not affect the actual implementation in practice. Then the new definition becomes "for each $i\in N\setminus \{\omega\}$, reporting both $p_i$ and $r_i$ truthfully is a dominant strategy". 

In the traditional settings, when $i$ misreports $p_i$, it will not affect the other players' participation/reports. However, in our model, if $i$ misreports $r_i$ (i.e., $i$ does not invite all her neighbours), some players may not be able to participate any more. For example, if a player $j$ can only be invited by $i$ and if $i$ does not invite $j$, then $j$ and all the other players connected to $j$ will not be able to join the mechanism. Therefore, misreporting in this model will affect the participation of the others, which is a key challenge in the design.

\section{Resource Allocation}
Resource allocation has been the mostly studied setting for mechanism design. Auctions such as VCG mechanism have been developed for resource allocation to achieve the desirable properties such as truthfulness, efficiency, and individual rationality. 
Another important property of resource allocation is maximizing the seller's revenue. It has been shown that VCG mechanism cannot maximize the revenue at the same time and especially, for combinatorial settings, the revenue of the seller may decrease when there are more buyers~\cite{DBLP:books/cu/NRTV2007,RevenueMonotonicity2011}. This also explains why under VCG, a buyer can create false ids to pretend to be multiple buyers to pay less for the same set of items \cite{yokoo2001robust}. To further maximize the seller's revenue, \citeauthor{myerson1981optimal} \cite{myerson1981optimal} proposed the very first mechanism to achieve the optimal revenue for selling a single item. The idea is to add a reserve price on top of the VCG and the reserve price requires buyers' valuation distributions. However, in practice, the most efficient way to increase the revenue is to seek more buyers to compete for the item. This is also justified in theory by \citeauthor{bulow1994auctions}~\cite{bulow1994auctions} and they showed that in the single-item setting, the optimal revenue with $n$ buyers is not more than the revenue of VCG with $n+1$ buyers. In practice, attracting one more buyer is much easier than finding the buyers' valuation distributions, and therefore the auctioneer prefers to spend more effort to attract more participants than optimising the mechanisms.

In order to seek more participants, traditionally, the seller would spend some cost via search engines or social media to promote the auction. However, the return of the promotion is unpredictable, and it is possible that the promotion does not bring any valuable participants. Under social networks, we could instead use participants' social connections to promote the auction. More importantly, we want to make sure the promotion will increase the seller's revenue. However, participants are competitors for the auctioned resources and they have no incentive to invite each other by default. Therefore, the challenge is that the auctioneer does not want to lose from the promotion and the participants do not want to do free promotions. This is a new design problem which contains both promotion and auction together and it has not been well studied in the literature. Recent studies discovered that this is a promising research direction worth further investigation \cite{DBLP:conf/ijcai/LiHZ20,li2018customer,li2019diffusion,10.5555/3398761.3398947,zhao2018selling,DBLP:conf/ecai/ZhangZZ20,li2017mechanism,DBLP:conf/aaai/KawasakiBTTY20}.

The problem can be framed as follows. A seller sells a set of items and the potential buyers are willing to buy the items (a combinatorial setting). Then a participant's type $p_i$ in this setting is a valuation function for all bundles of the items. 
The goal for the seller is to design a mechanism such that participants are incentivized to both report their valuation function truthfully and invite all their neighbours to join the sale. Also the seller's revenue should not be smaller than directly selling the items to her neighbours only (an incentive for the seller to apply the mechanism). 

Given the above setting, some novel mechanisms have recently been proposed for single-item settings and multiple identical items settings \cite{li2017mechanism, zhao2018selling}. The intuition behind their mechanisms is that a buyer can potentially benefit from inviting her neighbours. If a buyer does not invite anyone and wins the item, he will get a utility say $x$. If the buyer invited all her neighbours, then the buyer may lose the item, but the mechanism guarantees that the buyer's new utility $y$ is not less than $x$. This could be understood as that the buyer first buys the item and then resells it to her neighbours with a higher price. If the reselling price is not higher than the current buyer's valuation, then the buyer will keep the item. This also explains that why their mechanisms are not efficient (not maximizing social welfare). One important property of their mechanisms is that the seller's revenue is guaranteed to be non-decreasing, which means that the seller is incentivized to use the diffusion mechanisms to increase her revenue.

To get complete efficiency, VCG can be extended to incentivize buyers to invite each other, but the revenue of the seller will be negative \cite{li2017mechanism}. It is not hard to prove that under the network setting, it is impossible to achieve truthfulness, individual rationality and efficiency together with non-negative revenue for the seller (which can be proved by converting a simple model (where a seller connects to buyer A and buyer A further connects to buyer B) into a bilateral trading case~\cite{10.5555/3398761.3398947,DBLP:conf/ijcai/LiHZ20}). One interesting open question is how much efficiency we can approximate given that the seller's revenue is non-negative. We believe that this largely depends on the structure of the network and the buyers' valuation distributions. The mechanisms cited above, except for the VCG, has no guarantee in terms of approximating efficiency.  

Another more challenging open question is to design similar mechanisms for multiple heterogeneous items settings. It has been shown that the problem becomes extremely difficult when we move just one step forward from single-item to multiple homogeneous items cases~\cite{zhao2018selling,DBLP:conf/aaai/KawasakiBTTY20}. The difficulty comes from the fact that the allocation and price of one participant can be easily influenced by its invitees and siblings~\cite{zhao2018selling}. In order to get the property of truthfulness, we need to maximize everyone's utility, which leads to a super complex multiagent optimization problem. One possible breakthrough is to restrict the action space of the participants to simplify the optimization. Also there is a hierarchy problem following their invitations, where an inviter has a higher priority in the optimization than her invitees.

\section{Task Collaboration}
Task allocation is another important field of mechanism design. One setting is where there is one task to be finished and there are a number of workers who can perform the task with different (privately known) costs, and the goal is to find a suitable worker~\cite{shehory1998methods}. This setting can be transferred to the resource allocation setting (the task can be treated as a resource). What we consider here are cooperative games such as coalitional games and crowdsrourcing games. In a coalitional game, players can form groups to achieve different rewards and the goal is to design a reward distribution mechanism to incentivize them to work together as one group (a.k.a. grand coalition)~\cite{CooperativeGameTheory2011}. 
In a crowdsourcing game, we have a set of tasks to do and the goal is to find the best set of workers to perform them such as tasks on Amazon Mechanical Turk and Google Image Labeler~\cite{howe2006rise,deng2009imagenet}. Different from the resource allocation settings, we have both collaboration and competition here. The collaboration comes from the fact they need to work together to accomplish the tasks and the competition is due to the reward distribution.

In order to incentivize the players to collaborate in a coalitional game, the literature has focused on designing proper reward distribution mechanisms~\cite{CooperativeGameTheory2011}. Shapley value is one of the well-known distribution mechanisms satisfying many desirable properties~\cite{shapley1953value}. Core is another important property which says a distribution is a core if no subset of players can deviate from the grand coalition to receive better rewards~\cite{scarf1967core}. Shapley value is computable for all coalitional games, but core may not exist for all the games. Since a larger group of players work together can achieve more, we consider how to incentivize players to invite others (via their social connections) to join the game, which is widely applicable in practice. This is not well-considered in the literature and cannot be solved with the existing solutions.

Consider a simple example where initially only player $P_1$ is in the game and $P_1$ can achieve a utility $x$ alone. If $P_1$ has a neighbour $P_2$ whose ability is the same as $P_1$, i.e., they together cannot achieve more than $x$. If we apply Shapley value, $P_1$ receives $x$ without inviting $P_2$, but it is reduced to $x/2$ after inviting $P_2$. Thus, $P_1$ is not incentivized to invite $P_2$ if the reward is distributed by Shapley value. Actually, in this simple example, the reward distributed to $P_1$ should be at least $x$ in order to incentivize $P_1$ to invite others.

Therefore, our goal is to design new reward distribution mechanisms to incentivize them to not only work together, but also invite more players to cooperate together to receive better rewards. We assume that the players are connected to form a network and initially only a subset of them are in the game. The challenge is that we cannot treat inviters and invitees the same and in principle inviters have higher priorities/weights than invitees. The question is how to use this priority to define the reward distribution. 

\citeauthor{zhang2020collaborative}~\cite{zhang2020collaborative} studied a data acquisition setting by modifying Shapley value such that the join of invitees may only increase the Shapley value that the inviters can get before the invitation. By doing so, they group all players by layers, the first layer contains all the initial players, the second layer contains all the players directly invited by the first layer and so on. If we treat each layer as one group, then the reward distributed to the $k$-th layer is the marginal contribution when the $k$-th layer join after the first $k-1$ layers. However, the participation of the higher layers would not bring any benefit to the lower layers, so they proposed to take some portion of each layer's reward to be shared with its parent layers for the strong invitation incentive.

Incentivizing participants to invite others to join a challenging task has appeared in practice since the 2009 DARPA Network Challenge~\cite{tang2011reflecting}. It was a competition for first finding the red weather balloons at 10 previously undisclosed locations in the continental United States. A team from MIT won the competition and they proposed a novel reward distribution mechanism to attract more than 5000 participants to join the team via social platforms~\cite{pickard2011time}. The intuition of their solution is that the reward is not just given to the participants who find the balloons first, but also to the people who invited them (until the root of the invitation chain). In theory, their mechanism satisfies the invitation incentive we want to achieve here, which has been proved by \citeauthor{zhang2020incentives}~\cite{zhang2020incentives}. They further showed that any combination of weighted Shapley value and permission structure can give us invitation incentives and the solution from the winning team is a special example. One interesting open question is to further characterize all the mechanisms under coalitional games to satisfy invitation incentives.

As mentioned above, a player is incentivized to invite others because she will be rewarded for the invitation. This, however, can lead to another challenge where a player creates multiple fake nodes in the network to gain more benefit, which is known as false-name attack~\cite{yokoo2001robust}. It is easy to check that the MIT winning solution cannot prevent false-name attacks. \citeauthor{chen2013sybil}~\cite{chen2013sybil} proposed a solution for false-name proof in expectation in a single question answering scenario. For the deterministic setting, \citeauthor{ijcai2020-59}~\cite{ijcai2020-59} showed that requiring both invitation incentive and false-name proofness leaves us very limited design space. They also considered the manipulation of merging multiple nodes into one node. It is still open what mechanisms we can get under different combinations of all the desirable properties in coalitional games.

\section{Matching}
In both resource and task allocation, we often use utility transfer (monetary payment) to achieve the design goals. However, in some settings like matching, monetary payment is not an option. Matching is a well-studied field consisting of problems like house allocation, stable marriage problem and kidney exchange~\cite{hylland1979efficient,roth1985common,roth2004kidney,sonmez2020incentivized}.
In a matching setting, players have mainly ordinal preferences among different matchings, e.g. a player prefers house $a$ to house $b$ but does not have an exact value for each house. Also in kidney exchange, a donor cannot charge the patient who received the donation. Without the tool of utility transfer, the literature has focused on making optimal or stable matching such as the top trading cycle algorithm for house allocation and the Gale-Shapley algorithm for stable marriage problem~\cite{shapley1974cores,Gale1962College}.

We further study the matching problem from the perspective of attracting more participants. For example, in a house allocation or stable marriage problem, more participants will create better exchanges and make participants more satisfied. Again, in kidney exchange, if we have more donors and patients, we will be able to form more exchanges and benefit more patients~\cite{rothmaking, sonmez2020incentivized}.

The challenge we tackle here is how to incentivize existing participants to invite others who are not in the matching game yet. Existing solutions cannot solve this. Consider the top trading cycle algorithm for house exchange, if a participant $P_1$ invited another participant $P_2$, $P_2$ may compete with $P_1$ for the same house. For the Gale-Shapley algorithm in a stable marriage problem, if a man $M_1$ invited another man $M_2$ who has the same preference as $M_1$ does, then $M_2$ will compete with $M_1$ for the same women. Even if $M_1$ invites a woman $W_2$, $W_2$ could again invite $M_2$ to compete with $M_1$. Thus, both men and women would not invite others by default.

Therefore, we need to investigate new matching mechanisms to incentivize existing participants to invite new participants. Similar to the resource and task allocation, the key is that an inviters' match should not be sacrificed by their invitees. In traditional settings, we allow participants to have a full preference among all participants without any constraints, but in the network setting, we need to add constraints on their preferences in order to incentivize them to invite each other. For the example mentioned above, if $P_2$ competes with $P_1$ for the same house, assume their preferred house is with $P_3$ and $P_3$ does not know $P_2$, even if $P_3$ prefers $P_2$'s house to $P_1$'s house, we cannot allow $P_2$ and $P_3$ exchange directly (otherwise, $P_1$ would not invite $P_2$ to the game). The challenge here is how to interpret this kind of constraint in the matching process. \citeauthor{zhengyue2020}~\cite{zhengyue2020} proved that to incentivize participants to invite each other, we cannot further have the traditional optimality. Thus, the existing matching mechanisms cannot work in the new setting.

\section{Voting}
Voting is another important field in mechanism design where monetary transfer is not possible~\cite{nisan2007introduction,ComSOC2016}. Due to the development of social networks, more and more pools/votings are conducted online by inviting participants on social media~\cite{boldi2009voting,boldi2011viscous,escoffier2019convergence}. On one hand, we hope more participants can join a voting to make the results reflect the opinions of the majority. On the other hand, the invited participants are often the candidates' friends, which makes the results unfair/biased to some extent. To combat this challenge, we propose to design new voting rules such that more participants are invited by the existing participants even if they have different preferences. More importantly, we hope the voting results are not far from the results we can get when they all can participate without invitation. Intuitively, if it is a voting for choosing one winner, assume that the winner is $a$ when all participants vote under a classical voting rule. 
Then, under the new voting rule, except for attracting more voters, we also hope that the final voting result is not far from $a$. If such mechanism exists, we can always start a voting with a small number of voters in the community. The initial voters will spread the voting to their network and eventually all the voters in the community will be invited and the results still reflect the preferences of the majority.

The challenge here is that a voter preferring outcome $a$ to $b$ does not have incentives to invite voters preferring $b$ to $a$. This also explains why online votings are eventually a competition of the number of friends they have invited between the candidates. We need to change this by incentivizing the voters to invite all voters they know, even though they may have different preferences. To the best of our knowledge, no solution has been found for this.

One possible way to tackle the bias in online voting is liquid democracy (proxy voting). Voting via social network can easily realize the delegation process. Besides, those who are indifferent can transfer their voting right to knowledgeable people~\cite{kahng2018liquid,blum2016liquid}. However, existing mechanisms satisfying the delegation process cannot achieve a better outcome than each voter votes directly~\cite{caragiannis2019contribution}.

\section{Conclusion}
We have highlighted a new mechanism design challenge under the social network environment where each player is connected with some players (her neighbours) and the player does not know the others on the network. The design goal is to incentivize the players who are already in the game to further invite their neighbours to join the game, even though they are competing for the same resources, tasks or matches. We have emphasized four domains: auctions, coalitional games, matching and voting. They all have different goals and face different challenges, but in terms of incentivizing the players to invite each other, they share the same principle that invitees cannot sacrifice their inviters' utilities. We have seen some progress in this direction on auctions and coalitional games, but there are still many fundamental open questions to be answered. Of course, the diffusion study is not limited to the four domains. Any mechanism design settings where there is a need to attract more players can be studied under this framework.


%


\balance

\begin{thebibliography}{52}


\ifx \showCODEN    \undefined \def \showCODEN     #1{\unskip}     \fi
\ifx \showDOI      \undefined \def \showDOI       #1{#1}\fi
\ifx \showISBNx    \undefined \def \showISBNx     #1{\unskip}     \fi
\ifx \showISBNxiii \undefined \def \showISBNxiii  #1{\unskip}     \fi
\ifx \showISSN     \undefined \def \showISSN      #1{\unskip}     \fi
\ifx \showLCCN     \undefined \def \showLCCN      #1{\unskip}     \fi
\ifx \shownote     \undefined \def \shownote      #1{#1}          \fi
\ifx \showarticletitle \undefined \def \showarticletitle #1{#1}   \fi
\ifx \showURL      \undefined \def \showURL       {\relax}        \fi
\providecommand\bibfield[2]{#2}
\providecommand\bibinfo[2]{#2}
\providecommand\natexlab[1]{#1}
\providecommand\showeprint[2][]{arXiv:#2}

\bibitem[\protect\citeauthoryear{Adgate}{Adgate}{2020}]%
        {RDF_W3}
\bibfield{author}{\bibinfo{person}{Brad Adgate}.}
  \bibinfo{year}{2020}\natexlab{}.
\newblock \bibinfo{title}{In A First, Google Ad Revenue Expected To Drop In
  2020 Despite Growing Digital Ad Market}.
\newblock
  \bibinfo{howpublished}{\url{https://www.forbes.com/sites/bradadgate/2020/06/22/in-a-first-google-ad-revenue-expected-to-drop-in-2020-despite-growing-digital-ad-market/?sh=7c2a63af607d}}.
\newblock


\bibitem[\protect\citeauthoryear{Blum and Zuber}{Blum and Zuber}{2016}]%
        {blum2016liquid}
\bibfield{author}{\bibinfo{person}{Christian Blum} {and}
  \bibinfo{person}{Christina~Isabel Zuber}.} \bibinfo{year}{2016}\natexlab{}.
\newblock \showarticletitle{Liquid Democracy: Potentials, Problems, and
  Perspectives}.
\newblock \bibinfo{journal}{\emph{Journal of Political Philosophy}}
  \bibinfo{volume}{24}, \bibinfo{number}{2} (\bibinfo{year}{2016}),
  \bibinfo{pages}{162--182}.
\newblock


\bibitem[\protect\citeauthoryear{Boldi, Bonchi, Castillo, and Vigna}{Boldi
  et~al\mbox{.}}{2009}]%
        {boldi2009voting}
\bibfield{author}{\bibinfo{person}{Paolo Boldi}, \bibinfo{person}{Francesco
  Bonchi}, \bibinfo{person}{Carlos Castillo}, {and} \bibinfo{person}{Sebastiano
  Vigna}.} \bibinfo{year}{2009}\natexlab{}.
\newblock \showarticletitle{Voting in Social Networks}. In
  \bibinfo{booktitle}{\emph{Proceedings of the 18th ACM Conference on
  Information and Knowledge Management}}. \bibinfo{pages}{777--786}.
\newblock


\bibitem[\protect\citeauthoryear{Boldi, Bonchi, Castillo, and Vigna}{Boldi
  et~al\mbox{.}}{2011}]%
        {boldi2011viscous}
\bibfield{author}{\bibinfo{person}{Paolo Boldi}, \bibinfo{person}{Francesco
  Bonchi}, \bibinfo{person}{Carlos Castillo}, {and} \bibinfo{person}{Sebastiano
  Vigna}.} \bibinfo{year}{2011}\natexlab{}.
\newblock \showarticletitle{Viscous Democracy for Social Networks}.
\newblock \bibinfo{journal}{\emph{Commun. ACM}} \bibinfo{volume}{54},
  \bibinfo{number}{6} (\bibinfo{year}{2011}), \bibinfo{pages}{129--137}.
\newblock


\bibitem[\protect\citeauthoryear{Brandt, Conitzer, Endriss, Lang, and
  Procaccia}{Brandt et~al\mbox{.}}{2016}]%
        {ComSOC2016}
\bibfield{author}{\bibinfo{person}{Felix Brandt}, \bibinfo{person}{Vincent
  Conitzer}, \bibinfo{person}{Ulle Endriss}, \bibinfo{person}{J\'{e}r\^{o}me
  Lang}, {and} \bibinfo{person}{Ariel~D. Procaccia}.}
  \bibinfo{year}{2016}\natexlab{}.
\newblock \bibinfo{booktitle}{\emph{Handbook of Computational Social Choice}
  (\bibinfo{edition}{1st} ed.)}.
\newblock \bibinfo{publisher}{Cambridge University Press},
  \bibinfo{address}{USA}.
\newblock
\showISBNx{1107060435}


\bibitem[\protect\citeauthoryear{Bulow and Klemperer}{Bulow and
  Klemperer}{1994}]%
        {bulow1994auctions}
\bibfield{author}{\bibinfo{person}{Jeremy Bulow} {and} \bibinfo{person}{Paul
  Klemperer}.} \bibinfo{year}{1994}\natexlab{}.
\newblock \bibinfo{booktitle}{\emph{Auctions vs. negotiations}}.
\newblock \bibinfo{type}{{T}echnical {R}eport}. \bibinfo{institution}{National
  Bureau of Economic Research}.
\newblock


\bibitem[\protect\citeauthoryear{Caragiannis and Micha}{Caragiannis and
  Micha}{2019}]%
        {caragiannis2019contribution}
\bibfield{author}{\bibinfo{person}{Ioannis Caragiannis} {and}
  \bibinfo{person}{Evi Micha}.} \bibinfo{year}{2019}\natexlab{}.
\newblock \showarticletitle{A Contribution to the Critique of Liquid
  Democracy}. In \bibinfo{booktitle}{\emph{Proceedings of the Twenty-Eighth
  International Joint Conference on Artificial Intelligence, {IJCAI-19}}}.
  \bibinfo{publisher}{International Joint Conferences on Artificial
  Intelligence Organization}, \bibinfo{pages}{116--122}.
\newblock


\bibitem[\protect\citeauthoryear{Chalkiadakis, Elkind, and
  Wooldridge}{Chalkiadakis et~al\mbox{.}}{2011}]%
        {CooperativeGameTheory2011}
\bibfield{author}{\bibinfo{person}{Georgios Chalkiadakis},
  \bibinfo{person}{Edith Elkind}, {and} \bibinfo{person}{Michael Wooldridge}.}
  \bibinfo{year}{2011}\natexlab{}.
\newblock \bibinfo{booktitle}{\emph{Computational Aspects of Cooperative Game
  Theory (Synthesis Lectures on Artificial Inetlligence and Machine Learning)}
  (\bibinfo{edition}{1st} ed.)}.
\newblock \bibinfo{publisher}{Morgan \& Claypool Publishers}.
\newblock
\showISBNx{1608456528}


\bibitem[\protect\citeauthoryear{Chen, Wang, Yu, and Zhang}{Chen
  et~al\mbox{.}}{2013}]%
        {chen2013sybil}
\bibfield{author}{\bibinfo{person}{Wei Chen}, \bibinfo{person}{Yajun Wang},
  \bibinfo{person}{Dongxiao Yu}, {and} \bibinfo{person}{Li Zhang}.}
  \bibinfo{year}{2013}\natexlab{}.
\newblock \showarticletitle{Sybil-proof Mechanisms in Query Incentive
  Networks}. In \bibinfo{booktitle}{\emph{Proceedings of the fourteenth ACM
  conference on Electronic commerce}}. \bibinfo{pages}{197--214}.
\newblock


\bibitem[\protect\citeauthoryear{Clarke}{Clarke}{1971}]%
        {Clarke1971}
\bibfield{author}{\bibinfo{person}{Edward Clarke}.}
  \bibinfo{year}{1971}\natexlab{}.
\newblock \showarticletitle{Multipart pricing of public goods}.
\newblock \bibinfo{journal}{\emph{Public Choice}} \bibinfo{volume}{11},
  \bibinfo{number}{1} (\bibinfo{year}{1971}), \bibinfo{pages}{17--33}.
\newblock


\bibitem[\protect\citeauthoryear{Deng, Dong, Socher, Li, Li, and Li}{Deng
  et~al\mbox{.}}{2009}]%
        {deng2009imagenet}
\bibfield{author}{\bibinfo{person}{Jia Deng}, \bibinfo{person}{Wei Dong},
  \bibinfo{person}{Richard Socher}, \bibinfo{person}{Li{-}Jia Li},
  \bibinfo{person}{Kai Li}, {and} \bibinfo{person}{Fei{-}Fei Li}.}
  \bibinfo{year}{2009}\natexlab{}.
\newblock \showarticletitle{ImageNet: {A} Large-scale Hierarchical Image
  Database}. In \bibinfo{booktitle}{\emph{2009 {IEEE} Computer Society
  Conference on Computer Vision and Pattern Recognition {(CVPR} 2009), 20-25
  June 2009, Miami, Florida, {USA}}}. \bibinfo{publisher}{{IEEE} Computer
  Society}, \bibinfo{pages}{248--255}.
\newblock


\bibitem[\protect\citeauthoryear{Edelman, Ostrovsky, and Schwarz}{Edelman
  et~al\mbox{.}}{2007}]%
        {GSP2007}
\bibfield{author}{\bibinfo{person}{Benjamin Edelman}, \bibinfo{person}{Michael
  Ostrovsky}, {and} \bibinfo{person}{Michael Schwarz}.}
  \bibinfo{year}{2007}\natexlab{}.
\newblock \showarticletitle{{Internet Advertising and the Generalized
  Second-Price Auction: Selling Billions of Dollars Worth of Keywords}}.
\newblock \bibinfo{journal}{\emph{American Economic Review}}
  \bibinfo{volume}{97}, \bibinfo{number}{1} (\bibinfo{date}{March}
  \bibinfo{year}{2007}), \bibinfo{pages}{242--259}.
\newblock


\bibitem[\protect\citeauthoryear{Escoffier, Gilbert, and
  Pass-Lanneau}{Escoffier et~al\mbox{.}}{2019}]%
        {escoffier2019convergence}
\bibfield{author}{\bibinfo{person}{Bruno Escoffier}, \bibinfo{person}{Hugo
  Gilbert}, {and} \bibinfo{person}{Ad{\`e}le Pass-Lanneau}.}
  \bibinfo{year}{2019}\natexlab{}.
\newblock \showarticletitle{The Convergence of Iterative Delegations in Liquid
  Democracy in a Social Network}. In \bibinfo{booktitle}{\emph{International
  Symposium on Algorithmic Game Theory}}. Springer, \bibinfo{pages}{284--297}.
\newblock


\bibitem[\protect\citeauthoryear{Gale and Shapley}{Gale and Shapley}{1962}]%
        {Gale1962College}
\bibfield{author}{\bibinfo{person}{D. Gale} {and} \bibinfo{person}{L.~S.
  Shapley}.} \bibinfo{year}{1962}\natexlab{}.
\newblock \showarticletitle{College Admissions and the Stability of Marriage}.
\newblock \bibinfo{journal}{\emph{Amer. Math. Monthly}} \bibinfo{volume}{69},
  \bibinfo{number}{1} (\bibinfo{year}{1962}), \bibinfo{pages}{9--15}.
\newblock


\bibitem[\protect\citeauthoryear{Gibbard}{Gibbard}{1973}]%
        {gibbard1973manipulation}
\bibfield{author}{\bibinfo{person}{Allan Gibbard}.}
  \bibinfo{year}{1973}\natexlab{}.
\newblock \showarticletitle{Manipulation of Voting Schemes: A General Result}.
\newblock \bibinfo{journal}{\emph{Econometrica: Journal of the Econometric
  Society}} (\bibinfo{year}{1973}), \bibinfo{pages}{587--601}.
\newblock


\bibitem[\protect\citeauthoryear{Groves}{Groves}{1973}]%
        {Groves1973}
\bibfield{author}{\bibinfo{person}{Theodore Groves}.}
  \bibinfo{year}{1973}\natexlab{}.
\newblock \showarticletitle{Incentives in Teams}.
\newblock \bibinfo{journal}{\emph{Econometrica}} \bibinfo{volume}{41},
  \bibinfo{number}{4} (\bibinfo{year}{1973}), \bibinfo{pages}{617--31}.
\newblock


\bibitem[\protect\citeauthoryear{Howe}{Howe}{2006}]%
        {howe2006rise}
\bibfield{author}{\bibinfo{person}{Jeff Howe}.}
  \bibinfo{year}{2006}\natexlab{}.
\newblock \showarticletitle{The Rise of Crowdsourcing}.
\newblock \bibinfo{journal}{\emph{Wired Magazine}} \bibinfo{volume}{14},
  \bibinfo{number}{6} (\bibinfo{year}{2006}), \bibinfo{pages}{1--4}.
\newblock


\bibitem[\protect\citeauthoryear{Hylland and Zeckhauser}{Hylland and
  Zeckhauser}{1979}]%
        {hylland1979efficient}
\bibfield{author}{\bibinfo{person}{Aanund Hylland} {and}
  \bibinfo{person}{Richard Zeckhauser}.} \bibinfo{year}{1979}\natexlab{}.
\newblock \showarticletitle{The Efficient Allocation of Individuals to
  Positions}.
\newblock \bibinfo{journal}{\emph{Journal of Political Economy}}
  \bibinfo{volume}{87}, \bibinfo{number}{2} (\bibinfo{year}{1979}),
  \bibinfo{pages}{293--314}.
\newblock


\bibitem[\protect\citeauthoryear{Kahng, Mackenzie, and Procaccia}{Kahng
  et~al\mbox{.}}{2018}]%
        {kahng2018liquid}
\bibfield{author}{\bibinfo{person}{Anson Kahng}, \bibinfo{person}{Simon
  Mackenzie}, {and} \bibinfo{person}{Ariel~D Procaccia}.}
  \bibinfo{year}{2018}\natexlab{}.
\newblock \showarticletitle{Liquid Democracy: An Algorithmic Perspective}.
\newblock \bibinfo{journal}{\emph{AAAI 2018}} (\bibinfo{year}{2018}).
\newblock


\bibitem[\protect\citeauthoryear{Kawasaki, Barrot, Takanashi, Todo, and
  Yokoo}{Kawasaki et~al\mbox{.}}{2020}]%
        {DBLP:conf/aaai/KawasakiBTTY20}
\bibfield{author}{\bibinfo{person}{Takehiro Kawasaki},
  \bibinfo{person}{Nathana{\"{e}}l Barrot}, \bibinfo{person}{Seiji Takanashi},
  \bibinfo{person}{Taiki Todo}, {and} \bibinfo{person}{Makoto Yokoo}.}
  \bibinfo{year}{2020}\natexlab{}.
\newblock \showarticletitle{Strategy-Proof and Non-Wasteful Multi-Unit Auction
  via Social Network}. In \bibinfo{booktitle}{\emph{The Thirty-Fourth {AAAI}
  Conference on Artificial Intelligence, {AAAI} 2020, New York, NY, USA,
  February 7-12, 2020}}. \bibinfo{publisher}{{AAAI} Press},
  \bibinfo{pages}{2062--2069}.
\newblock


\bibitem[\protect\citeauthoryear{Li, Hao, and Zhao}{Li et~al\mbox{.}}{2020}]%
        {DBLP:conf/ijcai/LiHZ20}
\bibfield{author}{\bibinfo{person}{Bin Li}, \bibinfo{person}{Dong Hao}, {and}
  \bibinfo{person}{Dengji Zhao}.} \bibinfo{year}{2020}\natexlab{}.
\newblock \showarticletitle{Incentive-Compatible Diffusion Auctions}. In
  \bibinfo{booktitle}{\emph{Proceedings of the Twenty-Ninth International Joint
  Conference on Artificial Intelligence, {IJCAI} 2020}},
  \bibfield{editor}{\bibinfo{person}{Christian Bessiere}} (Ed.).
  \bibinfo{publisher}{ijcai.org}, \bibinfo{pages}{231--237}.
\newblock


\bibitem[\protect\citeauthoryear{Li, Hao, Zhao, and Yokoo}{Li
  et~al\mbox{.}}{2019}]%
        {li2019diffusion}
\bibfield{author}{\bibinfo{person}{Bin Li}, \bibinfo{person}{Dong Hao},
  \bibinfo{person}{Dengji Zhao}, {and} \bibinfo{person}{Makoto Yokoo}.}
  \bibinfo{year}{2019}\natexlab{}.
\newblock \showarticletitle{Diffusion and Auction on Graphs}. In
  \bibinfo{booktitle}{\emph{Proceedings of the Twenty-Eighth International
  Joint Conference on Artificial Intelligence, {IJCAI-19}}}.
  \bibinfo{publisher}{International Joint Conferences on Artificial
  Intelligence Organization}, \bibinfo{pages}{435--441}.
\newblock


\bibitem[\protect\citeauthoryear{Li, Hao, Zhao, and Zhou}{Li
  et~al\mbox{.}}{2017}]%
        {li2017mechanism}
\bibfield{author}{\bibinfo{person}{Bin Li}, \bibinfo{person}{Dong Hao},
  \bibinfo{person}{Dengji Zhao}, {and} \bibinfo{person}{Tao Zhou}.}
  \bibinfo{year}{2017}\natexlab{}.
\newblock \showarticletitle{Mechanism Design in Social Networks}. In
  \bibinfo{booktitle}{\emph{Proceedings of the Thirty-First AAAI Conference on
  Artificial Intelligence}} (San Francisco, California, USA).
  \bibinfo{publisher}{AAAI Press}, \bibinfo{pages}{586--592}.
\newblock


\bibitem[\protect\citeauthoryear{Li, Hao, Zhao, and Zhou}{Li
  et~al\mbox{.}}{2018}]%
        {li2018customer}
\bibfield{author}{\bibinfo{person}{Bin Li}, \bibinfo{person}{Dong Hao},
  \bibinfo{person}{Dengji Zhao}, {and} \bibinfo{person}{Tao Zhou}.}
  \bibinfo{year}{2018}\natexlab{}.
\newblock \showarticletitle{Customer Sharing in Economic Networks with Costs}.
  In \bibinfo{booktitle}{\emph{Proceedings of the Twenty-Seventh International
  Joint Conference on Artificial Intelligence, {IJCAI-18}}}.
  \bibinfo{publisher}{International Joint Conferences on Artificial
  Intelligence Organization}, \bibinfo{pages}{368--374}.
\newblock


\bibitem[\protect\citeauthoryear{MarketingDive}{MarketingDive}{2020}]%
        {TikTok}
\bibfield{author}{\bibinfo{person}{MarketingDive}.}
  \bibinfo{year}{2020}\natexlab{}.
\newblock \bibinfo{title}{TikTok unveils first shoppable livestream with
  Ntwrk}.
\newblock
  \bibinfo{howpublished}{\url{https://www.marketingdive.com/news/tiktok-unveils-first-shoppable-livestream-with-ntwrk/583992/}}.
\newblock


\bibitem[\protect\citeauthoryear{Myerson}{Myerson}{1981}]%
        {myerson1981optimal}
\bibfield{author}{\bibinfo{person}{Roger~B Myerson}.}
  \bibinfo{year}{1981}\natexlab{}.
\newblock \showarticletitle{Optimal Auction Design}.
\newblock \bibinfo{journal}{\emph{Mathematics of Operations Research}}
  \bibinfo{volume}{6}, \bibinfo{number}{1} (\bibinfo{year}{1981}),
  \bibinfo{pages}{58--73}.
\newblock


\bibitem[\protect\citeauthoryear{Myerson and Satterthwaite}{Myerson and
  Satterthwaite}{1983}]%
        {myerson1983efficient}
\bibfield{author}{\bibinfo{person}{Roger~B Myerson} {and}
  \bibinfo{person}{Mark~A Satterthwaite}.} \bibinfo{year}{1983}\natexlab{}.
\newblock \showarticletitle{Efficient Mechanisms for Bilateral Trading}.
\newblock \bibinfo{journal}{\emph{Journal of Economic Theory}}
  \bibinfo{volume}{29}, \bibinfo{number}{2} (\bibinfo{year}{1983}),
  \bibinfo{pages}{265--281}.
\newblock


\bibitem[\protect\citeauthoryear{Nisan, Roughgarden, Tardos, and
  Vazirani}{Nisan et~al\mbox{.}}{2007a}]%
        {nisan2007introduction}
\bibfield{editor}{\bibinfo{person}{Noam Nisan}, \bibinfo{person}{Tim
  Roughgarden}, \bibinfo{person}{{\'{E}}va Tardos}, {and}
  \bibinfo{person}{Vijay~V. Vazirani}} (Eds.).
  \bibinfo{year}{2007}\natexlab{a}.
\newblock \bibinfo{booktitle}{\emph{Introduction to Mechanism Design (for
  Computer Scientists)}}.
\newblock \bibinfo{publisher}{Cambridge University Press}. 209--242 pages.
\newblock


\bibitem[\protect\citeauthoryear{Nisan, Roughgarden, Tardos, and
  Vazirani}{Nisan et~al\mbox{.}}{2007b}]%
        {DBLP:books/cu/NRTV2007}
\bibfield{editor}{\bibinfo{person}{Noam Nisan}, \bibinfo{person}{Tim
  Roughgarden}, \bibinfo{person}{{\'{E}}va Tardos}, {and}
  \bibinfo{person}{Vijay~V. Vazirani}} (Eds.).
  \bibinfo{year}{2007}\natexlab{b}.
\newblock \bibinfo{booktitle}{\emph{Profit Maximization in Mechanism Design}}.
\newblock \bibinfo{publisher}{Cambridge University Press}. 331--361 pages.
\newblock


\bibitem[\protect\citeauthoryear{Perez}{Perez}{2020}]%
        {TikTok2}
\bibfield{author}{\bibinfo{person}{Sarah Perez}.}
  \bibinfo{year}{2020}\natexlab{}.
\newblock \bibinfo{title}{TikTok partners with Shopify on social commerce}.
\newblock
  \bibinfo{howpublished}{\url{https://techcrunch.com/2020/10/27/tiktok-invests-in-social-commerce-via-new-shopify-partnership/}}.
\newblock


\bibitem[\protect\citeauthoryear{Pickard, Pan, Rahwan, Cebrian, Crane, Madan,
  and Pentland}{Pickard et~al\mbox{.}}{2011}]%
        {pickard2011time}
\bibfield{author}{\bibinfo{person}{Galen Pickard}, \bibinfo{person}{Wei Pan},
  \bibinfo{person}{Iyad Rahwan}, \bibinfo{person}{Manuel Cebrian},
  \bibinfo{person}{Riley Crane}, \bibinfo{person}{Anmol Madan}, {and}
  \bibinfo{person}{Alex Pentland}.} \bibinfo{year}{2011}\natexlab{}.
\newblock \showarticletitle{Time-critical Social Mobilization}.
\newblock \bibinfo{journal}{\emph{Science}} \bibinfo{volume}{334},
  \bibinfo{number}{6055} (\bibinfo{year}{2011}), \bibinfo{pages}{509--512}.
\newblock


\bibitem[\protect\citeauthoryear{Rastegari, Condon, and Leyton-Brown}{Rastegari
  et~al\mbox{.}}{2011}]%
        {RevenueMonotonicity2011}
\bibfield{author}{\bibinfo{person}{Baharak Rastegari}, \bibinfo{person}{Anne
  Condon}, {and} \bibinfo{person}{Kevin Leyton-Brown}.}
  \bibinfo{year}{2011}\natexlab{}.
\newblock \showarticletitle{Revenue monotonicity in deterministic,
  dominant-strategy combinatorial auctions}.
\newblock \bibinfo{journal}{\emph{Artificial Intelligence}}
  \bibinfo{volume}{175}, \bibinfo{number}{2} (\bibinfo{year}{2011}),
  \bibinfo{pages}{441 -- 456}.
\newblock
\showISSN{0004-3702}
\urldef\tempurl%
\url{https://doi.org/10.1016/j.artint.2010.08.005}
\showDOI{\tempurl}


\bibitem[\protect\citeauthoryear{Roth}{Roth}{1985}]%
        {roth1985common}
\bibfield{author}{\bibinfo{person}{Alvin~E Roth}.}
  \bibinfo{year}{1985}\natexlab{}.
\newblock \showarticletitle{Common and conflicting interests in two-sided
  matching markets}.
\newblock \bibinfo{journal}{\emph{European Economic Review}}
  \bibinfo{volume}{27}, \bibinfo{number}{1} (\bibinfo{year}{1985}),
  \bibinfo{pages}{75--96}.
\newblock


\bibitem[\protect\citeauthoryear{Roth, S{\"o}nmez, and {\"U}nver}{Roth
  et~al\mbox{.}}{2004}]%
        {roth2004kidney}
\bibfield{author}{\bibinfo{person}{Alvin~E Roth}, \bibinfo{person}{Tayfun
  S{\"o}nmez}, {and} \bibinfo{person}{M~Utku {\"U}nver}.}
  \bibinfo{year}{2004}\natexlab{}.
\newblock \showarticletitle{Kidney exchange}.
\newblock \bibinfo{journal}{\emph{The Quarterly journal of economics}}
  \bibinfo{volume}{119}, \bibinfo{number}{2} (\bibinfo{year}{2004}),
  \bibinfo{pages}{457--488}.
\newblock


\bibitem[\protect\citeauthoryear{Roth and Shorrer}{Roth and Shorrer}{2018}]%
        {rothmaking}
\bibfield{author}{\bibinfo{person}{Benjamin~N Roth} {and}
  \bibinfo{person}{Ran~I Shorrer}.} \bibinfo{year}{2018}\natexlab{}.
\newblock \showarticletitle{Making Marketplaces Safe: Dominant Individual
  Rationality and Applications to Market Design}.
\newblock \bibinfo{howpublished}{\url{https://ssrn.com/abstract=3073027}}.
\newblock  (\bibinfo{year}{2018}).
\newblock


\bibitem[\protect\citeauthoryear{Satterthwaite}{Satterthwaite}{1975}]%
        {satterthwaite1975strategy}
\bibfield{author}{\bibinfo{person}{Mark~Allen Satterthwaite}.}
  \bibinfo{year}{1975}\natexlab{}.
\newblock \showarticletitle{Strategy-proofness and Arrow's Conditions:
  Existence and Correspondence Theorems for Voting Procedures and Social
  Welfare Functions}.
\newblock \bibinfo{journal}{\emph{Journal of Economic Theory}}
  \bibinfo{volume}{10}, \bibinfo{number}{2} (\bibinfo{year}{1975}),
  \bibinfo{pages}{187--217}.
\newblock


\bibitem[\protect\citeauthoryear{Scarf}{Scarf}{1967}]%
        {scarf1967core}
\bibfield{author}{\bibinfo{person}{Herbert~E Scarf}.}
  \bibinfo{year}{1967}\natexlab{}.
\newblock \showarticletitle{The Core of an N Person Game}.
\newblock \bibinfo{journal}{\emph{Econometrica: Journal of the Econometric
  Society}} (\bibinfo{year}{1967}), \bibinfo{pages}{50--69}.
\newblock


\bibitem[\protect\citeauthoryear{Shapley and Scarf}{Shapley and Scarf}{1974}]%
        {shapley1974cores}
\bibfield{author}{\bibinfo{person}{Lloyd Shapley} {and}
  \bibinfo{person}{Herbert Scarf}.} \bibinfo{year}{1974}\natexlab{}.
\newblock \showarticletitle{On Cores and Indivisibility}.
\newblock \bibinfo{journal}{\emph{Journal of Mathematical Economics}}
  \bibinfo{volume}{1}, \bibinfo{number}{1} (\bibinfo{year}{1974}),
  \bibinfo{pages}{23--37}.
\newblock


\bibitem[\protect\citeauthoryear{Shapley}{Shapley}{1953}]%
        {shapley1953value}
\bibfield{author}{\bibinfo{person}{Lloyd~S Shapley}.}
  \bibinfo{year}{1953}\natexlab{}.
\newblock \showarticletitle{A Value for n-person Games}.
\newblock \bibinfo{journal}{\emph{Contributions to the Theory of Games}}
  \bibinfo{volume}{2}, \bibinfo{number}{28} (\bibinfo{year}{1953}),
  \bibinfo{pages}{307--317}.
\newblock


\bibitem[\protect\citeauthoryear{Shehory and Kraus}{Shehory and Kraus}{1998}]%
        {shehory1998methods}
\bibfield{author}{\bibinfo{person}{Onn Shehory} {and} \bibinfo{person}{Sarit
  Kraus}.} \bibinfo{year}{1998}\natexlab{}.
\newblock \showarticletitle{Methods for Task Allocation via Agent Coalition
  Formation}.
\newblock \bibinfo{journal}{\emph{Artificial intelligence}}
  \bibinfo{volume}{101}, \bibinfo{number}{1-2} (\bibinfo{year}{1998}),
  \bibinfo{pages}{165--200}.
\newblock


\bibitem[\protect\citeauthoryear{Shubik}{Shubik}{2004}]%
        {shubik2004theory}
\bibfield{author}{\bibinfo{person}{Martin Shubik}.}
  \bibinfo{year}{2004}\natexlab{}.
\newblock \bibinfo{booktitle}{\emph{The Theory of Money and Financial
  Institutions}}. Vol.~\bibinfo{volume}{1}.
\newblock \bibinfo{publisher}{Mit Press}.
\newblock


\bibitem[\protect\citeauthoryear{S{\"o}nmez, {\"U}nver, and Yenmez}{S{\"o}nmez
  et~al\mbox{.}}{2020}]%
        {sonmez2020incentivized}
\bibfield{author}{\bibinfo{person}{Tayfun S{\"o}nmez}, \bibinfo{person}{M~Utku
  {\"U}nver}, {and} \bibinfo{person}{M~Bumin Yenmez}.}
  \bibinfo{year}{2020}\natexlab{}.
\newblock \showarticletitle{Incentivized kidney exchange}.
\newblock \bibinfo{journal}{\emph{American Economic Review}}
  \bibinfo{volume}{110}, \bibinfo{number}{7} (\bibinfo{year}{2020}),
  \bibinfo{pages}{2198--2224}.
\newblock


\bibitem[\protect\citeauthoryear{Tang, Cebrian, Giacobe, Kim, Kim, and
  Wickert}{Tang et~al\mbox{.}}{2011}]%
        {tang2011reflecting}
\bibfield{author}{\bibinfo{person}{John~C Tang}, \bibinfo{person}{Manuel
  Cebrian}, \bibinfo{person}{Nicklaus~A Giacobe}, \bibinfo{person}{Hyun-Woo
  Kim}, \bibinfo{person}{Taemie Kim}, {and} \bibinfo{person}{Douglas~"Beaker"
  Wickert}.} \bibinfo{year}{2011}\natexlab{}.
\newblock \showarticletitle{Reflecting on the DARPA Red Balloon Challenge}.
\newblock \bibinfo{journal}{\emph{Commun. ACM}} \bibinfo{volume}{54},
  \bibinfo{number}{4} (\bibinfo{year}{2011}), \bibinfo{pages}{78--85}.
\newblock


\bibitem[\protect\citeauthoryear{Vickrey}{Vickrey}{1961}]%
        {vickrey1961counterspeculation}
\bibfield{author}{\bibinfo{person}{William Vickrey}.}
  \bibinfo{year}{1961}\natexlab{}.
\newblock \showarticletitle{Counterspeculation, Auctions, and Competitive
  Sealed Tenders}.
\newblock \bibinfo{journal}{\emph{The Journal of Finance}}
  \bibinfo{volume}{16}, \bibinfo{number}{1} (\bibinfo{year}{1961}),
  \bibinfo{pages}{8--37}.
\newblock


\bibitem[\protect\citeauthoryear{Yokoo, Sakurai, and Matsubara}{Yokoo
  et~al\mbox{.}}{2001}]%
        {yokoo2001robust}
\bibfield{author}{\bibinfo{person}{Makoto Yokoo}, \bibinfo{person}{Yuko
  Sakurai}, {and} \bibinfo{person}{Shigeo Matsubara}.}
  \bibinfo{year}{2001}\natexlab{}.
\newblock \showarticletitle{Robust combinatorial auction protocol against
  false-name bids}.
\newblock \bibinfo{journal}{\emph{Artificial Intelligence}}
  \bibinfo{volume}{130}, \bibinfo{number}{2} (\bibinfo{year}{2001}),
  \bibinfo{pages}{167--181}.
\newblock


\bibitem[\protect\citeauthoryear{Zhang, Zhang, and Zhao}{Zhang
  et~al\mbox{.}}{2020a}]%
        {zhang2020collaborative}
\bibfield{author}{\bibinfo{person}{Wen Zhang}, \bibinfo{person}{Yao Zhang},
  {and} \bibinfo{person}{Dengji Zhao}.} \bibinfo{year}{2020}\natexlab{a}.
\newblock \showarticletitle{Collaborative Data Acquisition}. In
  \bibinfo{booktitle}{\emph{Proceedings of the 19th International Conference on
  Autonomous Agents and MultiAgent Systems}}. \bibinfo{pages}{1629--1637}.
\newblock


\bibitem[\protect\citeauthoryear{Zhang, Zhao, and Chen}{Zhang
  et~al\mbox{.}}{2020c}]%
        {10.5555/3398761.3398947}
\bibfield{author}{\bibinfo{person}{Wen Zhang}, \bibinfo{person}{Dengji Zhao},
  {and} \bibinfo{person}{Hanyu Chen}.} \bibinfo{year}{2020}\natexlab{c}.
\newblock \showarticletitle{Redistribution Mechanism on Networks}.
  \bibinfo{publisher}{International Foundation for Autonomous Agents and
  Multiagent Systems}, \bibinfo{address}{Richland, SC},
  \bibinfo{pages}{1620--1628}.
\newblock
\showISBNx{9781450375184}


\bibitem[\protect\citeauthoryear{Zhang, Zhao, and Zhang}{Zhang
  et~al\mbox{.}}{2020d}]%
        {DBLP:conf/ecai/ZhangZZ20}
\bibfield{author}{\bibinfo{person}{Wen Zhang}, \bibinfo{person}{Dengji Zhao},
  {and} \bibinfo{person}{Yao Zhang}.} \bibinfo{year}{2020}\natexlab{d}.
\newblock \showarticletitle{Incentivize Diffusion with Fair Rewards}. In
  \bibinfo{booktitle}{\emph{{ECAI} 2020 - 24th European Conference on
  Artificial Intelligence}} \emph{(\bibinfo{series}{Frontiers in Artificial
  Intelligence and Applications}, Vol.~\bibinfo{volume}{325})}.
  \bibinfo{publisher}{{IOS} Press}, \bibinfo{pages}{251--258}.
\newblock


\bibitem[\protect\citeauthoryear{Zhang, Zhang, and Zhao}{Zhang
  et~al\mbox{.}}{2020b}]%
        {ijcai2020-59}
\bibfield{author}{\bibinfo{person}{Yao Zhang}, \bibinfo{person}{Xiuzhen Zhang},
  {and} \bibinfo{person}{Dengji Zhao}.} \bibinfo{year}{2020}\natexlab{b}.
\newblock \showarticletitle{Sybil-proof Answer Querying Mechanism}. In
  \bibinfo{booktitle}{\emph{Proceedings of the Twenty-Ninth International Joint
  Conference on Artificial Intelligence, {IJCAI-20}}}.
  \bibinfo{publisher}{International Joint Conferences on Artificial
  Intelligence Organization}, \bibinfo{pages}{422--428}.
\newblock


\bibitem[\protect\citeauthoryear{Zhang and Zhao}{Zhang and Zhao}{2020}]%
        {zhang2020incentives}
\bibfield{author}{\bibinfo{person}{Yao Zhang} {and} \bibinfo{person}{Dengji
  Zhao}.} \bibinfo{year}{2020}\natexlab{}.
\newblock \showarticletitle{Incentives to Form Larger Coalitions when Players
  Have the Power to Choose}.
\newblock \bibinfo{journal}{\emph{CoRR}}  \bibinfo{volume}{abs/2011.09049}
  (\bibinfo{year}{2020}).
\newblock
\showeprint[arxiv]{2011.09049}


\bibitem[\protect\citeauthoryear{Zhao, Li, Xu, Hao, and Jennings}{Zhao
  et~al\mbox{.}}{2018}]%
        {zhao2018selling}
\bibfield{author}{\bibinfo{person}{Dengji Zhao}, \bibinfo{person}{Bin Li},
  \bibinfo{person}{Junping Xu}, \bibinfo{person}{Dong Hao}, {and}
  \bibinfo{person}{Nicholas~R. Jennings}.} \bibinfo{year}{2018}\natexlab{}.
\newblock \showarticletitle{Selling Multiple Items via Social Networks}. In
  \bibinfo{booktitle}{\emph{Proceedings of the 17th International Conference on
  Autonomous Agents and MultiAgent Systems, {AAMAS} 2018, Stockholm, Sweden,
  July 10-15, 2018}}. \bibinfo{publisher}{International Foundation for
  Autonomous Agents and Multiagent Systems Richland, SC, {USA} / {ACM}},
  \bibinfo{pages}{68--76}.
\newblock


\bibitem[\protect\citeauthoryear{Zheng, Yang, Zhang, and Zhao}{Zheng
  et~al\mbox{.}}{2020}]%
        {zhengyue2020}
\bibfield{author}{\bibinfo{person}{Yue Zheng}, \bibinfo{person}{Tianyi Yang},
  \bibinfo{person}{Wen Zhang}, {and} \bibinfo{person}{Dengji Zhao}.}
  \bibinfo{year}{2020}\natexlab{}.
\newblock \showarticletitle{Barter Exchange via Friends' Friends}.
\newblock \bibinfo{journal}{\emph{CoRR}}  \bibinfo{volume}{abs/2010.04933}
  (\bibinfo{year}{2020}).
\newblock
\showeprint[arxiv]{2010.04933}


\end{thebibliography}



\end{document}